\documentclass[journal=jacsat,manuscript=article]{achemso}

\usepackage[version=3]{mhchem} % Formula subscripts using \ce{}

\author{Uma Namangalam}
\affiliation{Department of Physics \& CAMOST, IISER Tirupati $-$ 517619, Andhra Pradesh, India.}
\author{Salvi Mohandas}
\affiliation{Present Address: Institut f\"ur Physik, Johannes Gutenberg-Universit\"at Mainz, Germany.}
\author{Hemanth Dinesan}
\affiliation{Present Address$:$ Laboratoire de Physique des Lasers CNRS in the Universit\'e Paris 13, France.}
\author{\\Sunil Kumar S.}
\email{sunil.phys@gmail.com}
\affiliation
{Department of Physics \& CAMOST, IISER Tirupati $-$ 517619, Andhra Pradesh, India.}

\title[An \textsf{achemso} demo]
  {Suppression of collision-induced dissociation in a supersonically expanding gas}

\abbreviations{IR,NMR,UV}
\keywords{American Chemical Society, \LaTeX}

\begin{document}


\begin{abstract}
 In high-resolution mass spectrometry, an electrospray ionization source is often paired with an ion-funnel to enhance ion transmission. Although it is established that ions experience collision-induced dissociation as they pass through this device, the impact of gas-flow dynamics on ion fragmentation remains unexplored. The present work demonstrates that the gas-flow dynamics from the capillary interface of an electrospray ionization source into an ion-funnel significantly reduces ion fragmentation. This reduction stems from the substantial decrease in the rate of increase in the internal energy of the ions resulting from the collisions with a supersonically expanding gas. The results of this study have significant consequences for systems that employ electrospray mass spectrometry and ion-mobility spectrometry.
\end{abstract}

\section{Introduction}
Electrospray ionization (ESI) is an ion generation technique extensively employed in mass spectrometry due to its versatility in producing multiply charged biomolecular ions in the gas phase with minimal fragmentation~\cite{Fenn1989-gh,Kebarle2009-bs}. The ions generated in an ESI source are commonly sampled by an ion transfer capillary (ITC), which serves as an interface between the atmospheric pressure region and a low-vacuum region. The ions are then transmitted through a skimmer or ion-funnel interface, which couples the low-vacuum stage to a high-vacuum stage of a mass spectrometer or other specialized instruments. While the ion-funnel interface is far superior to the skimmer interface in transferring the ions exiting the ITC (nearly 100 \% versus less than 10 \% \cite{Belov2000-hq}), a significant drawback is that the ions undergo fragmentation due to ion-gas collisions within the ion-funnel \cite{Shaffer1997-bv,Kelly2010-pg}. Since the ion-funnels work under relatively low-vacuum conditions (0.1-10 mbar), there will be several hundreds of collisions before the ions reach the next vacuum stage. Factors such as the radiofrequency (RF) potential and the DC gradients applied across the ion-funnel influence the extent of fragmentation \cite{Shaffer1998-yd,Kim2000-dx}. 

In an ESI source, the ions are conveyed through the ion transfer capillary via gas flow resulting from the pressure difference between the electrospray region and the first vacuum stage. The capillaries utilized for this interface typically feature small inner diameters in the range of a few hundred micrometers. This configuration results in a supersonic gas expansion from the ITC into the ion-funnel. The gas dynamics under supersonic expansion has been extensively investigated by several research groups \cite{Tejeda1996-zb,Gabelica2005-zc}. Also, the ion fragmentation occurring at the orifice$-$skimmer interface of an electrospray spectrometer has been examined by Schneider and Chen \cite{Schneider2000-gq}. However, the impact of such fluid flow on ion fragmentation within an ion-funnel interface has not been studied. In the current investigation, we present valuable insights into the effects of supersonic gas expansion on the fragmentation dynamics of gas-phase biomolecular ions in an ESI$-$ion-funnel interface.

\section{Experimental details}
\begin{figure}[h!]
\centering
\includegraphics[scale=0.58]{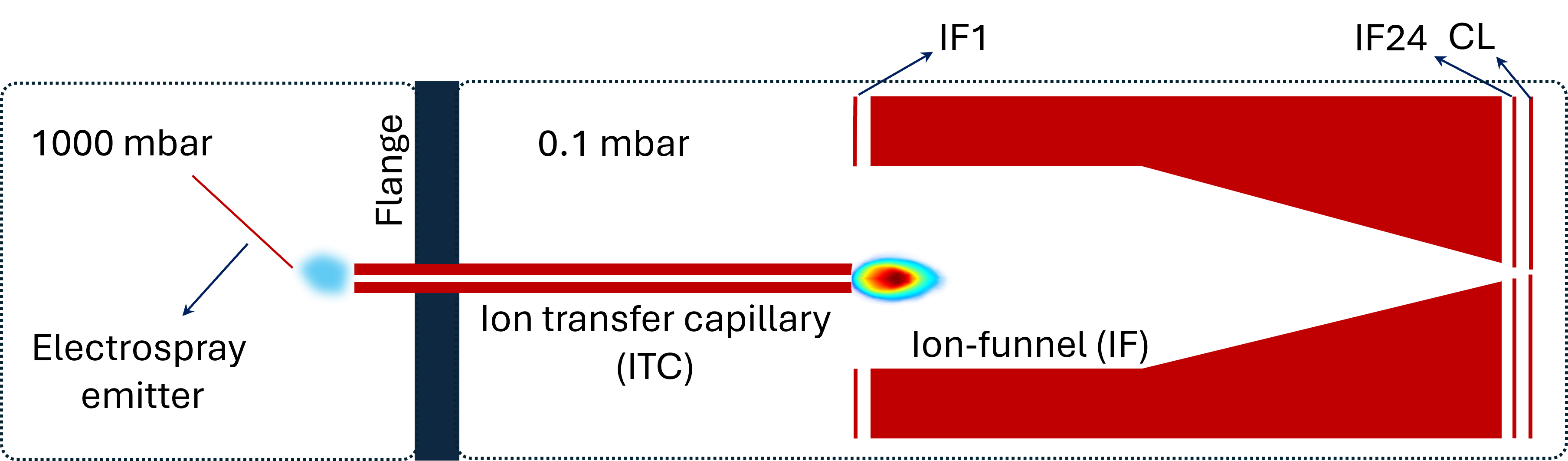}
\caption{\label{fig:esiif} Schematic of the experimental setup showing the atmospheric pressure region of the ESI source and the low-vacuum region of the ion-funnel. The electrospray emitter is oriented at an angle with respect to the ITC to avoid the straight jet of the sample. The ion-funnel electrodes between the first and the last electrodes are shown as a shaded block for simplicity. A schematic of the spray and the supersonic gas expansion from the capillary are also depicted.}
\end{figure}

We employed a home-built experimental setup that features an ESI source integrated with an ITC and an ion-funnel~\cite{Salvi2023-sg} (Fig. S1). A schematic of these two stages, most relevant to the current study, is presented in Fig.~\ref{fig:esiif}. The ion-funnel consists of 24 ring electrodes on which RF potentials are applied (with opposite polarity on alternate electrodes) for radial confinement of ions. A DC gradient is established across the ion-funnel by applying potentials on the first and last RF electrodes (IF1 and IF24, respectively) through a resistive chain for ion extraction along the axis. The ion-funnel is located in a vacuum chamber maintained at a pressure of 0.1 mbar. The conductance limiter (CL) is a ring electrode between the chamber housing the ion-funnel and the following high-vacuum stage, which houses a quadrupole ion guide (QIG). A DC potential is applied to the CL to focus the ions into the QIG. The exit of the ITC and the first electrode of the ion-funnel, IF1, are on the same plane by design.

While the molecular ions are transported through the ion-funnel, they undergo multiple collisions with the background gas, leading to their fragmentation in some cases. This collision-induced dissociation (CID) occurs when the collision energy is converted into the internal energy of the molecular ions so that they are vibrationally excited, and the internal energy exceeds the CID threshold \cite{Gabelica2005-zc}. The parent and fragment ions so generated are transported through the QIG, followed by a home-built quadrupole mass spectrometer. These ions are then guided through a 16-pole ion trap, which is used in the ion-guiding configuration for the present experiments, before they are counted by a microchannel plate-based detection system. 

\section{Methodology}
We used deprotonated 2’-deoxyadenosine 5’-monophosphate (d-dAMP) as a model system for the current study because the fragmentation characteristics of this molecular ion have been extensively investigated in the literature~\cite{ho1997studies,Marcum2009-bk,Kumar2011-qo}. The mass spectra analysis revealed that the DC potentials applied on the ion-funnel electrodes and the ITC significantly influence ion fragmentation. The extent of fragmentation was found to correlate with the strength of the DC gradient that guides the ions through the ion-funnel. To identify the key parameters influencing ion fragmentation and hence minimize it, we employed two schemes to control the DC gradient within the ion-funnel:
\begin{enumerate}
    \item \emph{GradIF}: Vary the potential on the first electrode of the ion-funnel (IF1, see Fig.~1) while keeping all other DC potentials on the ion-funnel electrodes constant ($V_{IF24} = -5$\,V and $V_{CL} = -1$\,V). In this case, the potential on the ITC is also adjusted such that $|V_{ITC}-V_{IF1}|=50\,$V. We refer to this configuration as \emph{GradIF}, denoting the potential difference $|V_{IF1}-V_{IF24}|$ as a measure of the potential gradient.
    \item \emph{GradITC}: Vary the potential of the ITC while keeping the DC potentials on the ion-funnel electrodes constant. We refer to this setting as \emph{GradITC}, where the potential difference $|V_{ITC}-V_{IF1}|$ is denoted as a gradient. In this case, the potential on IF1 is fixed at $-50\,$V. Note that \emph{GradIF}45 is equivalent to \emph{GradITC}50.
\end{enumerate}

\begin{figure}[h!]
\centering
\includegraphics[scale=0.6]{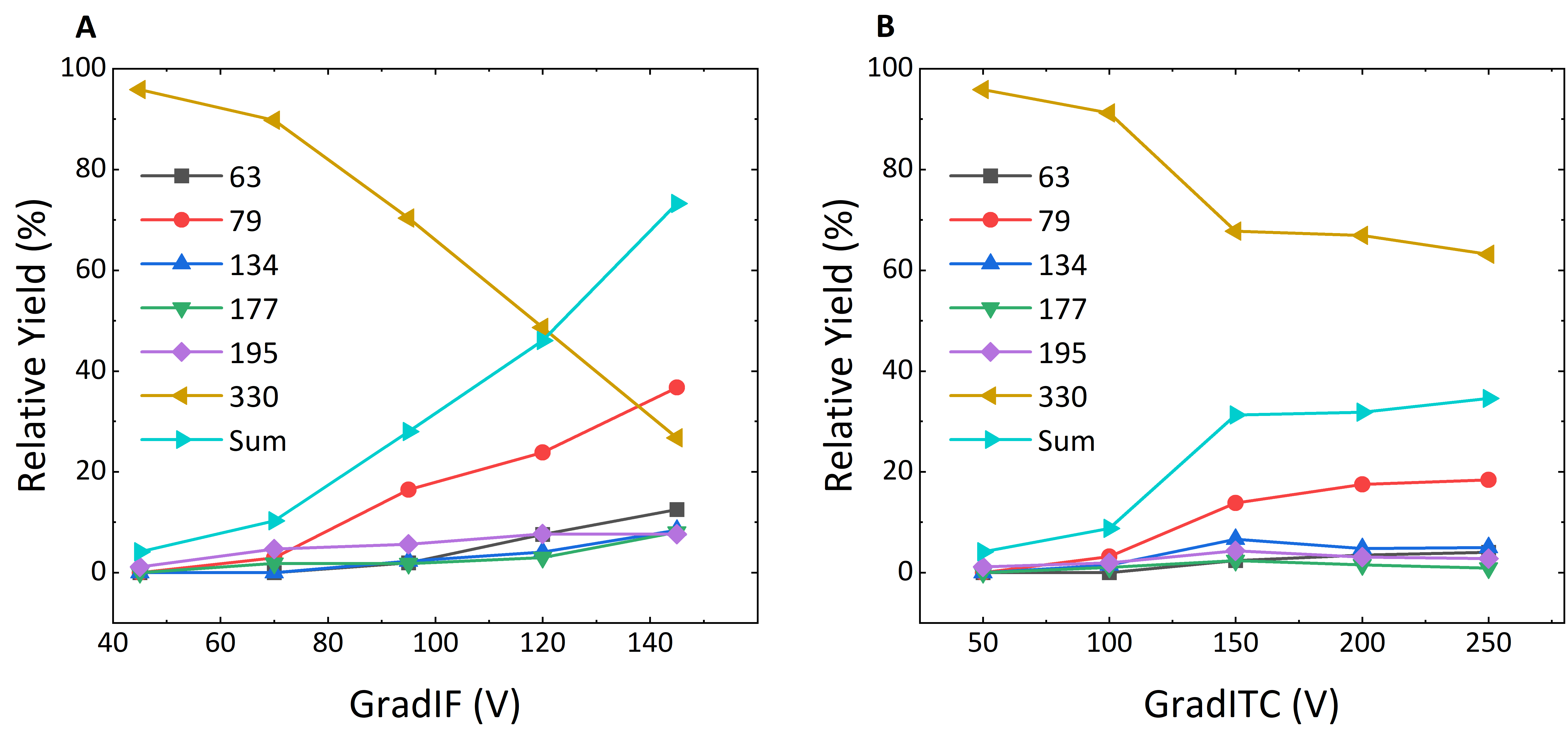}
\caption{\label{fig:cidddamp} CID pattern of d-dAMP as a function of DC gradient under schemes \emph{GradIF} (A) and \emph{GradITC} (B). The legends indicate $m/z$ values, and ``Sum" indicates the sum of relative yields. The extent of CID is higher for \emph{GradIF}.}
\end{figure}

The mass spectrum of d-dAMP consisted of ions of $m/z$: 330 (parent ion), 195 (loss of adenine base), 177 (loss of water from $m/z$ 195), 134 (deprotonated adenine), 79 ($\rm PO_3^-$), and 63 ($\rm PO_2^-$) (Figs. S2 and S3). The observed fragmentation spectrum is consistent with previously reported CID studies ~\cite{habibi1995ion,ho1997studies}. The mechanisms behind the generation of the major fragment ions have been well-established~\cite{ho1997studies,Kumar2011-qo}. Fig.~\ref{fig:cidddamp} shows the fragmentation pattern of d-dAMP under the configurations \emph{GradIF} and \emph{GradITC} with the ion-funnel operated under the RF setting of 90\,Vpp at 700\,kHz. The relative yields of parent and fragment ions formed during collisions across the ion-funnel were computed by determining the ratio of the areas under each mass peak to the total area of all the significant mass peaks. In both \emph{GradIF} and \emph{GradITC} configurations, the yield of the parent ion decreases, and that of the fragment ions increases as the gradient increases. 

\begin{figure}[ht]
\centering
\includegraphics[scale=0.55]{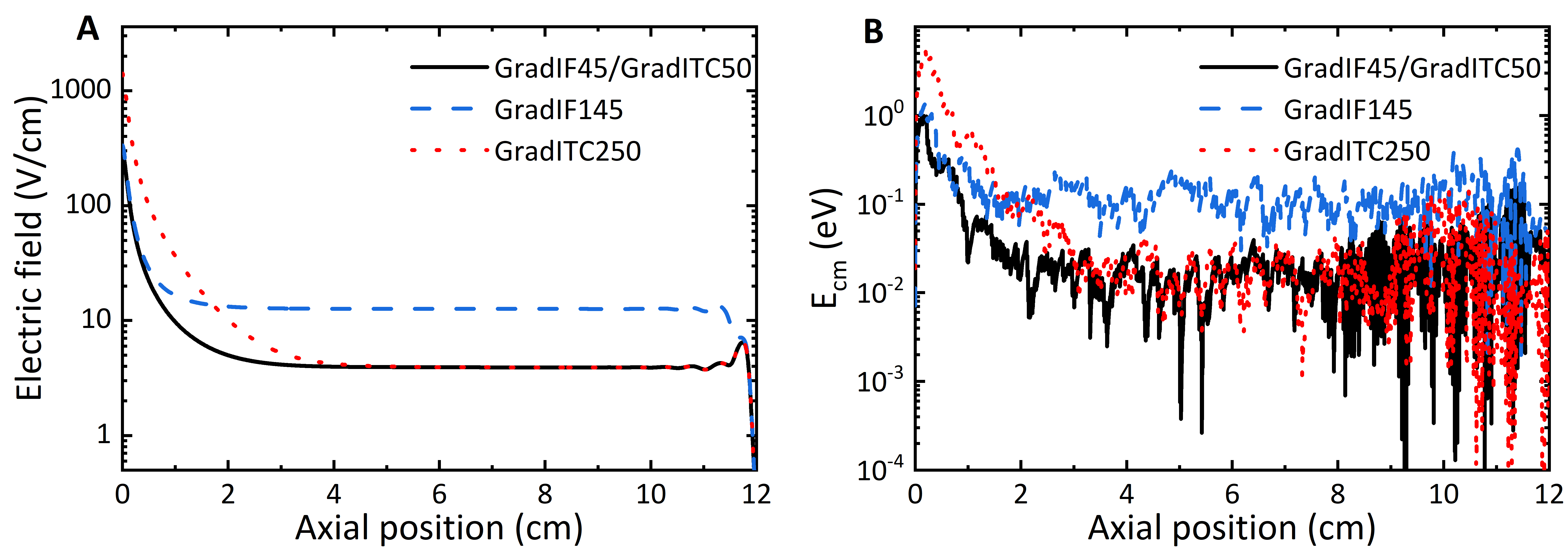}
\caption{\label{fig:efifecm}  Center of mass collision energy dependence on the electric field inside the ion funnel. {The abscissa label ``Axial position" refers to the position along the axis of the ion funnel with reference to the capillary exit.} (A) The electric field along the axis of the ion-funnel with the lowest and highest gradient settings on the ion-funnel under schemes \emph{GradITC} and \emph{GradIF}. Note that \emph{GradIF} = 45\,V and \emph{GradITC} = 50\,V are equivalent configurations. (B) The center of mass collision energy of a single ion for each collision under the above gradient settings.}
\end{figure}

Although the extent of fragmentation is anticipated to increase as the DC gradient increases, the difference in fragmentation rates under the two gradient settings is not straightforward. To understand the observed fragmentation trend, we plot in Fig.~\ref{fig:efifecm}A the electric field along the axis of the ion-funnel that results from the lowest and the highest gradient settings under \emph{GradIF} and \emph{GradITC}. The value of the electric field is the highest near the entrance of the ion-funnel with \emph{GradITC}250 where $V_{ITC} = -300$\,V and $V_{IF1} = -50$\,V. With this setting, since $V_{IF24}=-5$\,V, the electric field deep within the ion-funnel coincides with the lowest \emph{GradIF} setting: \emph{GradIF}45 where $V_{ITC}=-100$\,V and $V_{IF1}=-50$\,V. Note that the electric field deeper inside the ion-funnel for higher gradient settings under \emph{GradIF} is higher than that results from the highest gradient applied under \emph{GradITC}. Therefore, the ions experience the maximum acceleration with the highest \emph{GradITC} configuration close to the ion-funnel entrance. Under higher \emph{GradIF} configurations, the ions will undergo a higher acceleration deeper within the ion-funnel compared to that of \emph{GradITC} in the same region.

Now, referring back to Fig.~\ref{fig:cidddamp}, the extent of fragmentation is lower for \emph{GradITC}, despite the fact that the electric field in the entrance region of the ion-funnel is much higher than that in the \emph{GradIF} case. However, one might argue that, in the \emph{GradITC} case, the field with even the highest ITC potential is lower deep inside the ion-funnel, and the ion fragmentation could be higher with higher \emph{GradIF} settings in this region. This argument does not adequately explain the observations, as we illustrate below. Since the scope of this article is to comprehend the physics behind the difference in fragmentation characteristics under the two settings, we refrain from discussing the detailed differences in relative abundances of various fragment ions.

To comprehend the behavior of the fragmentation process within the ion-funnel, one needs to characterize the increase in internal energy of the ions due to collisions under the electric field settings used in the experiments. Toward this goal, we conducted numerical simulations on the ion transmission through the ion-funnel using SIMION ion optics software \cite{Appelhans2005-ki}. SIMION simulates ion trajectories through electric fields generated by the potentials (both DC and RF) applied on various electrode configurations. It allows for the incorporation of collisions of ions with gas molecules, based on a hard-sphere collision model to simulate elastic collisions. In our simulations, the gas molecules ($N_2$) were described by a Maxwell-Boltzmann velocity distribution at a temperature of 298\,K. We utilized a collision cross-section of $1.74\times10^{-18}\,\rm m^2$~\cite{zheng2017structural}, to characterize the interactions between d-dAMP and $N_2$. We simulated ion trajectories by modeling the ion-funnel, including a short segment of the ion transfer capillary (22.5\,mm long, 250\,$\mu$m inner diameter and 1.6\,mm outer diameter), to realistically characterize the electric field generated within the ion-funnel (Fig. S4). The exact RF and DC potentials employed in the experiments were used in the simulations. Since the ions exiting the capillary in an experiment would have velocities similar to those of the gas flow carrying the ions (corresponding to a kinetic energy of slightly less than an eV), the ions were initialized with an energy of 1\,eV starting from the exit of the ITC, and the energy of the ions soon after each collision was recorded (within an accuracy of $10\,$ns, the time step used for simulations). We verified that this energy is almost identical to the energy just before collision (Fig. S7). The time step was carefully chosen to be much shorter than the average collision time of around 100\,ns corresponding to a number density of $2.4\times10^{21}\,\rm{m^{-3}}$. The maximum amount of collisional energy that can be converted into the ion's internal energy is given by the total energy available in the center of mass system of the ion and the gas molecule (excluding the translational energy of the center of mass itself)~\cite{Shukla1993-at}. This is equal to $E_{cm}$ given by

\begin{eqnarray}
    E_{cm}=\frac{1}{2}m_{ion}u_{ion}^2\left(1+\frac{m_{ion}}{m_{gas}}\right)\label{eq:ecm}
\end{eqnarray}
where $m_{gas}$ and $m_{ion}$ are the masses of the gas molecule and the ion, respectively, and $u_{ion}$ is the ion velocity in the center of the mass frame ($u_{ion}=v_{ion}-v_{cm}$, $v_{ion}$ and $v_{cm}$, being the respective velocities of the ion and the center of mass in the lab frame). Fig.~\ref{fig:efifecm}B shows plots of the center of mass collision energy of an ion calculated for every collision occurring during the ion motion across the ion-funnel for the same gradient settings used for plotting the electric field in Fig.~\ref{fig:efifecm}A. In the case of \emph{GradIF}, an increase in gradient has only a slight influence on the $E_{cm}$ near the entrance of the ion-funnel, whereas its value deeper within the ion-funnel is much higher. In the case \emph{GradITC}, $E_{cm}$ increases significantly, though only near the entrance of the ion-funnel. It is striking to notice that $E_{cm}$ follows the overall trend in the electric field gradient. 

The increase in internal energy of the molecular ion due to a single collision may be expressed as \cite{Drahos2001-pc}

\begin{eqnarray}
    \Delta E_{int}=\eta' E_{cm}\label{eq:eta}
\end{eqnarray}
where $\eta'$ ($0\le\eta'\le1$) is the inelasticity parameter, typically much smaller than one (Section S-IV). For collision-induced fragmentation to occur, the internal energy of the ions should exceed the CID threshold. The lowest CID threshold for d-dAMP has been reported to be 3.74 eV, corresponding to the formation of $m/z$ 195 \cite{ho1997studies}. None of the single collisions resulted in $E_{cm}$ values greater than the threshold for even the extreme gradient settings under \emph{GradIF}, indicating that single collisions could not have led to fragmentation of the ions within the ion-funnel under these configurations (Fig.~\ref{fig:efifecm}B). In the case of \emph{GradITC}, some collisions close to the capillary exit has $E_{cm}$ values exceeding the dissociation thresholds of most channels at high gradients. However, it is unlikely that single collisions could have generated fragment ions in the case of \emph{GradITC} with high gradients, since such collisions must be highly inelastic. Therefore, we infer that the fragment ions we observed in our experiments mostly should have formed due to multiple collisions \cite{Blom1983-om,Mayer2009-he}. 

To describe the observed fragmentation behavior within a multiple collision model, we computed the cumulative values of $E_{cm}$ due to every collision that occurs during the transport of an ion through the ion-funnel. The cumulative $E_{cm}$ is then scaled with a parameter $\eta$, which we refer to as \emph{effective single-collision inelasticity parameter} to determine the total increase in the internal energy of the ions: $\Delta E_{int}^{tot}=\eta \sum E_{cm}$, where the summation runs over the number of collisions. Here, we assume that the value of $\eta$ does not change with collision energy, since the center of mass collision energy varies only over a small range, and any energy dependence on $\eta$ is neglected.

\begin{figure}[h!]
\centering
\includegraphics[scale=0.55]{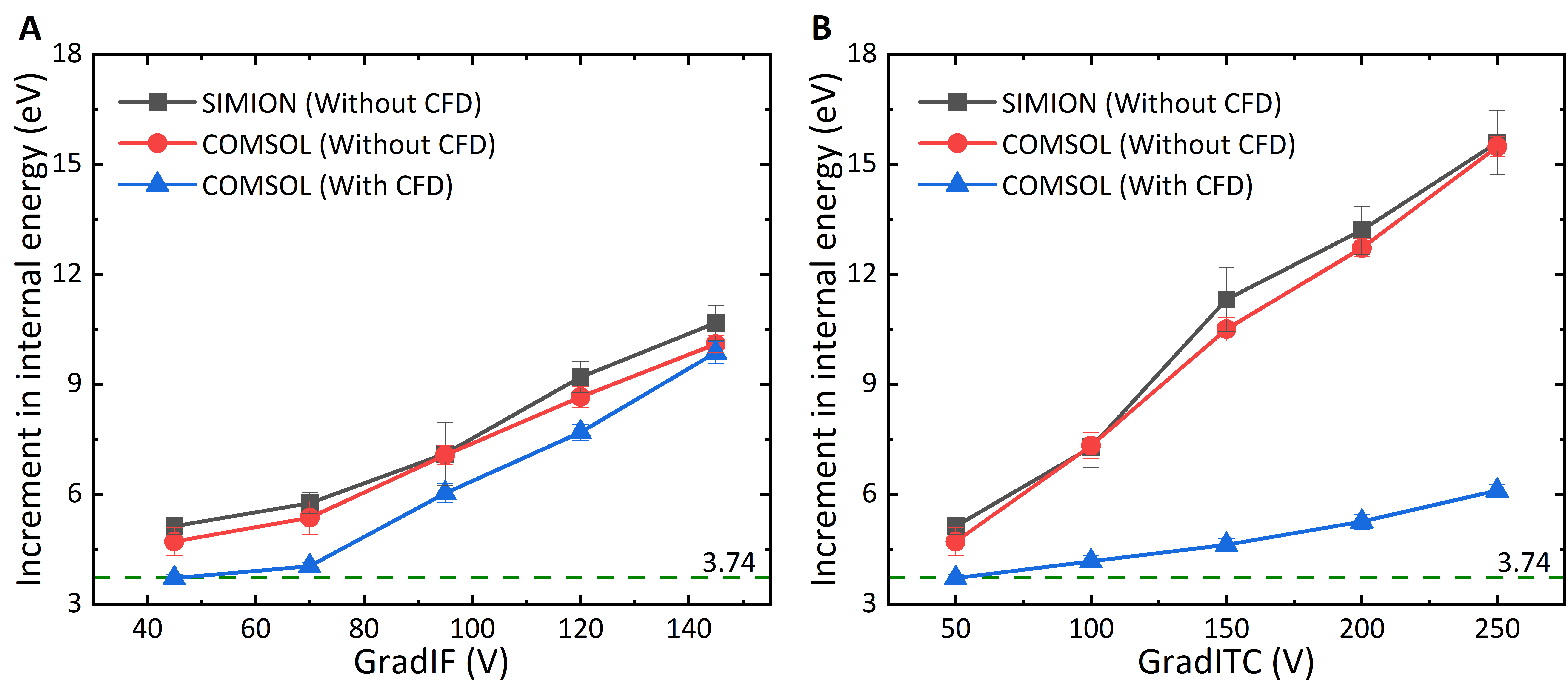}
\caption{\label{fig:eint} Increase in internal energy of ions computed by considering multiple collisions. (A) Case: \emph{GradIF} and (B) Case: \emph{GradITC}. The black squares are the results from SIMION simulations. The red dots and blue triangles correspond to COMSOL simulations without and with fluid flow, respectively. The horizontal line represents the lowest CID threshold of 3.74\,eV.}
\end{figure}

To find the value of $\eta$, we calibrated our fragmentation data using the CID thresholds reported by Ho and Kebarle \cite{ho1997studies,Kumar2011-qo}. We used the yield of $PO_3^-$ ($m/z:\,79$), the fragment ion of the highest abundance in our measurements, formed by binary dissociation of d-dAMP \cite{ho1997studies,Kumar2011-qo}. The data plotted in Fig.~\ref{fig:cidddamp}A for $m/z$ 79 was fitted with the equation:
\begin{eqnarray}
    Y(E)=Y_0\frac{(E-E_0)^n}{E}\label{eq:yield}
\end{eqnarray}
following the procedure proposed by Loh et al.~\cite{Loh1989-ui}. Here, $E$ is the collision energy and $E_0$ is the dissociation threshold. $E_0$, $Y_0$, and $n$ are free fitting parameters. This formula has been shown to describe well the threshold behavior of numerous ion-molecule collision cross-sections, especially when the ionic species is large enough to be characterized by numerous energy levels~\cite{Loh1989-ui}, which is indeed true for our analysis. In our experiment, $E$ represents the \emph{GradIF} value. The value of $E_0$ was found to be $60\pm10\,$V (Fig. S5), which should correspond to the dissociation threshold. Therefore, the $GradIF$ value was chosen to be equal to $E_0$ to perform simulations using SIMION to estimate $E_{cm}$ for each collision occurring across the ion-funnel using Eq.~(\ref{eq:ecm}). When multiplied by the effective inelasticity parameter $\eta$, the sum of all such values should correspond to the dissociation threshold for $PO_3^-$ (4.06\,eV)~\cite{ho1997studies}. Thus, the value of $\eta$ was determined to be $0.11\pm0.03$, where the error in $\eta$ was obtained by propagating the errors from the standard deviation of $\eta$ values found by performing ion trajectory simulations with five ions for five different \emph{GradIF} values chosen to fall within the error estimate of $E_0$. The same value of $\eta$ was used to compute the internal energy increase for different ion-funnel settings since this value should depend only on the collision system (d-dAMP with gas molecules), ignoring any energy dependence of $\eta$. The increments in internal energy of the ions so obtained are depicted in Fig.~\ref{fig:eint} with the square symbols at different DC gradients under \emph{GradIF} and \emph{GradITC} configurations. These simulations yielded an increase in the internal energy of the ions with an increase in the DC gradient across the ion-funnel in both cases. Notably, the rate of increase is significantly higher for \emph{GradITC}. This would imply that a more extensive fragmentation would occur in this case, contrary to the experimental findings.

\begin{figure}[h!]
\centering
\includegraphics[scale=0.70]{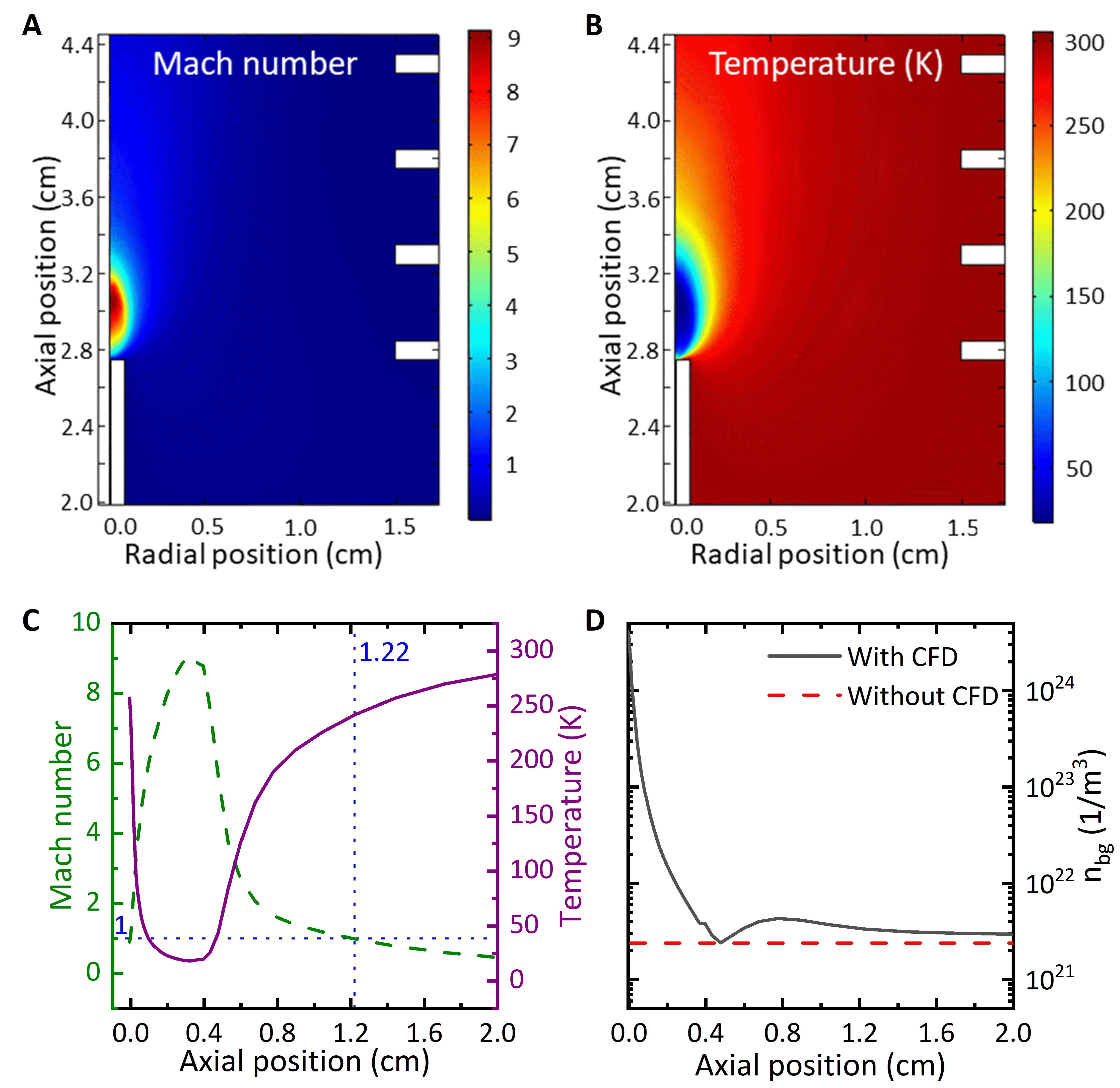}
\caption{\label{fig:fsimmt} Results from COMSOL simulations for the high Mach number flow. (A) Surface plot of the Mach number, (B) surface plot of temperature, (C) {line plots of Mach number and temperature} and (D) the number density variation along the axis of the ion-funnel. The surface plots additionally display segments of the ITC and IF electrodes. Note that the axial positions depicted in the top panels correspond to the simulation-defined positions, whereas those in the bottom panels are referenced from the capillary exit.}
\end{figure}

These observations led us to investigate the effect of fluid flow in dictating ion fragmentation within the ion-funnel. To account for this, we extended the simulations by modeling our experimental setup with the atmospheric pressure region, the capillary, and the first vacuum chamber, including the ion-funnel, the schematic of which is shown in Fig.~\ref{fig:esiif}. The gas-flow dynamics from the atmospheric pressure region of the ESI source (the source is not included in the simulations) to the low-vacuum region of the ion-funnel interface was simulated using the computational fluid dynamics (CFD) module for high Mach number flow implemented in COMSOL Multiphysics, a software package that allows one to simulate various physical phenomena such as fluid dynamics and ion dynamics in electric and magnetic fields~\cite{comsol-yk}. The Reynolds averaged Navier-Stokes equation was used to describe the turbulence characteristics of the gas flow. The computational details are available in Section S-V. The fluid flow features a supersonic expansion from the capillary exit, as shown in Fig.~\ref{fig:fsimmt}. The Mach zone (where Mach number = gas flow speed / sound speed $>1$) extends to about 12.2\,mm (Fig.~\ref{fig:fsimmt}C) from the capillary exit, which is less than the value obtained using the equation $d=(2/3)D\sqrt{p_i/p_o}=16.7\,\rm{mm}$~\cite{Fenn2000-xe,Morse1996-os}, applied for supersonic expansion through an aperture of diameter $D$, separating two regions of pressure $p_i$ and $p_o\ (<<p_i)$. The discrepancy arises because supersonic expansion results from a capillary and not from an orifice (Fig. S9). Further, the gas temperature reduces to below 50 K within the Mach zone (Fig.~\ref{fig:fsimmt}C). We also simulated the ion trajectory using COMSOL without and with CFD. COMSOL simulations yielded $\eta=0.16\pm0.03$ without CFD and $0.063\pm0.006$ with CFD. The increments in internal energy computed using COMSOL without and with CFD are presented in Fig.~\ref{fig:eint} using circles and triangles, respectively. The COMSOL results obtained without CFD match closely with those from SIMION simulations. Strikingly, the increase in internal energy predicted by the calculations with CFD is much smaller for \emph{GradITC} than for \emph{GradIF}. These results indicate that fragmentation with the \emph{GradITC} scheme must result in less fragmentation than in the \emph{GradIF} scheme, which now aligns with our experimental findings. These results emphasize the importance of incorporating gas-flow dynamics to comprehend ion fragmentation within an ion-funnel interface and similar scenarios.
\section{Results and discussion}
\begin{figure}[h!]
\centering
\includegraphics[scale=0.6]{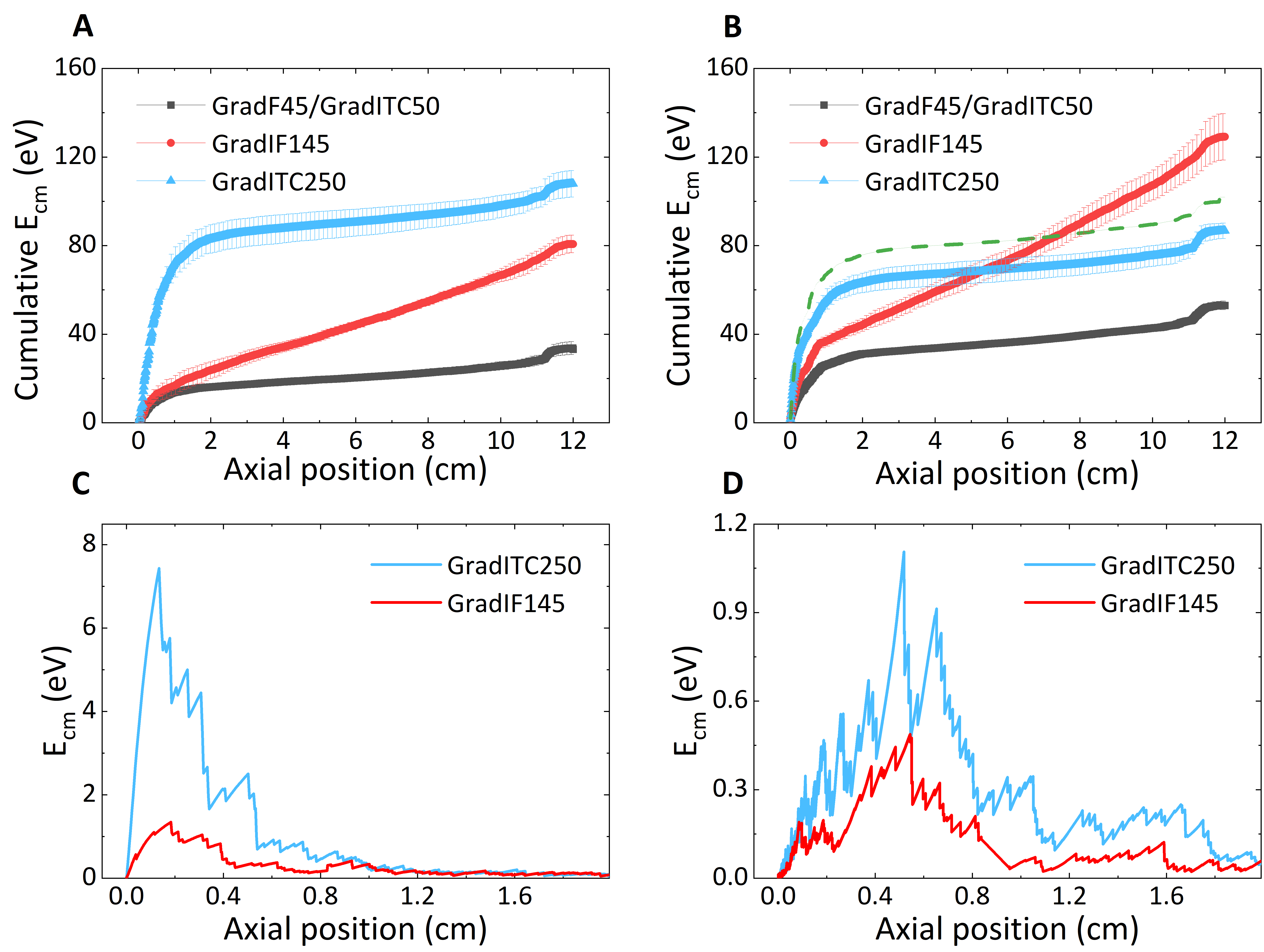}
\caption{\label{fig:comp_ecm}Cumulative $E_{cm}$ (A and B) and $E_{cm}$ for each collision for a single ion (C and D) as a function of distance along the axis of the ion-funnel (``Axial position"). Cumulative $E_{cm}$ is computed using data from five ions with standard deviations as error bars. (A) and (B) show results from simulations without and with CFD, respectively, except for the dashed line in (B), which is the result of the simulations done without CFD at a fixed temperature 300\,K, but with number density obtained from the CFD simulations. Note that the (C) and (D) show only 2\,cm downstream the capillary exit.}
\end{figure}

One might wonder exactly what causes the reduction in the internal energy of the ions with CFD. There are two possibilities: (A) the number of collisions is likely to be very different because of the gas flow dynamics within the Mach zone, which in turn could reduce the overall internal energy, and (B) the collision of the molecular ions with \emph{colder gas molecules} could also lower the internal energy. To address these questions, we examined the cumulative $E_{cm}$ for a single ion computed across the ion-funnel under the extreme gradient settings discussed in Fig.~\ref{fig:efifecm} without and with CFD as illustrated in the top panel of Fig.~\ref{fig:comp_ecm}. 
A slight deviation of the cumulative $E_{cm}$ from the monotonous trend toward the exit of the ion-funnel arises from the transient trapping of ions close to this region under the ion-funnel RF settings considered for this study. The cumulative $E_{cm}$, without CFD (Fig.~\ref{fig:comp_ecm}A), is the highest for \emph{GradITC}$=250$\,V, by the time the ion reaches the exit of the ion-funnel. The fast rise in the cumulative $E_{cm}$ with this setting near the capillary exit stems from the stronger electric field in this region (cf. Fig.~\ref{fig:efifecm}A). With CFD (Fig.~\ref{fig:comp_ecm}B), where the number density of the ions in the vicinity of the capillary exit is around three orders of magnitude higher (Fig.~\ref{fig:fsimmt}D), the scenario changes significantly. Here, the huge number of collisions with the background gas reduces the rate at which the cumulative $E_{cm}$ increases for \emph{GradITC}$=250$\,V, because the ions do not get sufficient time to gain high energy from the electric field near the capillary exit. Interestingly, for \emph{GradIF}$=145$\,V (also for \emph{GradITC}$=50$\,V), the cumulative $E_{cm}$ grows at a rate higher than the non-CFD case near the exit of the capillary despite having numerous collisions in this region. This seemingly counterintuitive behavior becomes easier to comprehend by considering the plots on the bottom panel of Fig.~\ref{fig:comp_ecm}, where $E_{cm}$ for every collision encountered by a single ion is plotted as a function of distance within 2\,cm from the capillary exit without (Fig.~\ref{fig:comp_ecm}C) and with (Fig.~\ref{fig:comp_ecm}D) CFD. For \emph{GradITC}$=250$\,V, the suppression of $E_{cm}$ is much stronger (a factor of six to seven) compared to that for \emph{GradIF}$=145$\,V (a factor of two to three). Beyond the Mach zone, the number density of the gas is more or less the same with and without CFD (Fig.~\ref{fig:fsimmt}D). Since the electric field for the \emph{GradIF} case is much higher in much of this region compared to that of \emph{GradITC}, the cumulative $E_{cm}$ increases substantially over the length of the ion-funnel in the former case leading to a much higher increase in internal energy. This analysis demonstrates that the fragmentation of molecular ions can transpire across the entirety of the ion-funnel. 

To address the question whether the gas temperature has any influence on the internal energy, we investigated the variation of $E_{cm}$ under a hypothetical scenario: the transfer of an ion through the ion-funnel was simulated without CFD at 300\,K, but with the number density obtained from the CFD simulation (dashed line in Fig.~\ref{fig:comp_ecm}B). Evidently, the cumulative $E_{cm}$ is higher in this case than that resulting from CFD, which implies that collisions with cold gas molecules indeed lowers the internal energy of the ions. Fig.~\ref{fig:fsimmt}C depicts the region where the temperature of the gas drops, and this is the region where the cumulative $E_{cm}$ decreases significantly. Based on this analysis, we infer that the decrease in the ions' internal energy at higher \emph{GradITC} configurations results from the dramatic increase in the number of ``cold" collisions with the gas molecules within the Mach zone.

It is worth mentioning that the quantitative results of our analysis can depend on the geometry of the setup, such as the length and the diameter of the capillary and the relative positioning of the capillary with respect to the ion-funnel. It will also depend on the pressure gradient across the capillary and the potentials on the ion-funnel. As a guideline for mass spectrometry systems, one could consider designing them to make use of the natural ``cooling" provided by the supersonic expansion and, hence, minimize the fragmentation that could result within an ion-funnel or a similar device.

In conclusion, this work demonstrates the significance of incorporating gas flow dynamics for accurately describing the observed fragmentation pattern in an electrospray$-$ion-funnel interface. A high potential difference between the ion transfer capillary and the ion-funnel does not necessarily lead to the fragmentation of ions because the frequent collisions of ions with supersonically expanding gas reduce the rate at which the internal energy of the ions increases due to the collisions. However, ion fragmentation can occur within the ion-funnel outside the Mach zone, because the continuous acceleration of the ions across the ion-funnel combined with ion-molecule collisions can still excite the ions slowly through their vibrational degrees of freedom, ultimately leading to their fragmentation once the dissociation threshold is crossed. These findings illustrate the importance of considering gas-flow dynamics in the design of mass spectrometers and ion-mobility spectrometers.


\section{Associated contents}
\paragraph{Supporting information:}
Supporting information available for  ``Suppression of collision-induced dissociation in a supersonically expanding gas"

\section{Author Information}
\paragraph*{Author contributions:}
Conceptualization: S.K.S. and U.N., Investigation: U.N. and S.K.S., Software (automation using LabVIEW) support: S.M., and H.D., Execution of the experiments and simulations, writing (original draft): S.K.S. and U.N., Review and editing: U.N., S.M, and H.D., Building and optimization of the experimental setup: S.M., U.N., and S.K.S., Supervision: S.K.S.
\begin{acknowledgement}

We extend our thanks to Abheek Roy and the other group members for the insightful discussions that ultimately led to the conclusion.

\end{acknowledgement}
\bibliography{CID_JACS}

\end{document}